\documentclass[12pt,a4paper]{article}

\usepackage{epsfig}
\usepackage{amsmath,amsfonts,amssymb}
\usepackage{cite}
\usepackage{colortbl}
\usepackage{pifont}

\providecommand{\openone}{\leavevmode\hbox{\small1\kern-3.8pt\normalsize1}}
\newcommand{\oh}{{\textstyle \scriptstyle \frac{1}{2}}}
\newcommand{\moh}{{\textstyle \scriptstyle - \frac{1}{2}}}

\newcommand{\mo}{{\textstyle \scriptscriptstyle -}1}

\newcommand{\RE}{\operatorname{Re}}
\newcommand{\IM}{\operatorname{Im}}

\newcommand{\vl}{V_L}
\newcommand{\vr}{V_R}
\newcommand{\gl}{g_L}
\newcommand{\gr}{g_R}

\begin{document}

\begin{center}
\begin{Large}
{\bf The fully differential top decay distribution}
\end{Large}

\vspace{0.5cm}
J.~A.~Aguilar--Saavedra$^{a}$, J. Boudreau$^{b}$, C. Escobar$^{c}$, J. Mueller$^{b}$ \\[1mm]
\begin{small}
{$^a$ Departamento de F\'{\i}sica Te\'orica y del Cosmos, 
Universidad de Granada, \\ E-18071 Granada, Spain} \\ 
{$^b$ Department of Physics and Astronomy, University of Pittsburgh, \\ Pittsburgh PA, United States of America} \\
{$^c$ Instituto de F\'{\i}sica Corpuscular, CSIC--Universitat de Val\`encia, E-46890 Paterna, Spain }
\end{small}
\end{center}

\begin{abstract}
We write down the four-dimensional fully differential decay distribution for the top quark decay $t \to Wb \to \ell \nu b$. We discuss how its eight physical parameters can be measured, either with a global fit or with the use of selected one-dimensional distributions and asymmetries. We give expressions for the top decay amplitudes for a general $tbW$ interaction, and show how the untangled measurement of the two components of the fraction of longitudinal $W$ bosons --- those with $b$ quark helicities of $1/2$ and $-1/2$, respectively ---  could improve the precision of a global fit to the $tbW$ vertex.
\end{abstract}

\section{Introduction}

The detailed study of the properties of the top quark has a widespread interest as a probe of new physics beyond the Standard Model (SM)~\cite{Bernreuther:2008ju,Han:2008xb,Aguilar-Saavedra:2014kpa}. In hadron collisions top quarks are produced in pairs, mediated by the QCD interaction, and they are also singly produced via electroweak interactions --- setting aside other sub-dominant production mechanisms in association with additional bosons $\gamma$, $W$, $Z$ or $H$. In contrast, the decay of the top (anti-)quarks is almost completely dominated by the mode $t \to W b$, with the subsequent decays of the $W$ bosons into charged leptons, $W \to \ell \nu$, $\ell = e,\mu,\tau$, or into quarks, $W \to q \bar q'$.

The top decay differential distribution in $t \to Wb \to \ell \nu$ is determined by four angles. The first two angles $(\theta,\phi)$ are the spherical coordinates of the $W$ boson momentum in the top quark rest frame, in some arbitrary reference system $(x,y,z)$. The remaining two angles $(\theta^*,\phi^*)$ are the spherical coordinates of the charged lepton momentum in the $W$ boson rest frame, in the reference system $(x',y',z')$ obtained by a ``standard'' boost from the previous one, such that the $\hat z'$ axis is in the direction $(\theta,\phi)$ of the $W$ boson momentum and the $\hat y'$ axis is in the $xy$ plane (see for example Ref.~\cite{libro} for a detailed discussion). One-dimensional distributions or asymmetries were already proposed long ago to measure some physical quantities involved in the top decay, such as the $W$ helicity fractions, measured from the $\theta^*$ distribution, or the top polarisation, measured from the $\theta$ and $\phi$ distributions~\cite{Kane:1991bg}. Since then, $W$ helicity fractions have been experimentally measured in a number of experiments, with the most precise measurements obtained by the ATLAS~\cite{Aaboud:2016hsq} and CMS~\cite{Khachatryan:2016fky} Collaborations, and the top polarisation in certain directions has also been measured in single top~\cite{Khachatryan:2015dzz,ATLAS:2016dcn} and top pair~\cite{Aad:2013ksa,Khachatryan:2016xws} production.  Other angular distributions, like the polar angle of the charged lepton in the top quark rest frame~\cite{Kuhn:1981md}, the azimuthal angle~\cite{Rindani:2011pk}, and the azimuthal angle in the laboratory frame~\cite{Godbole:2010kr}, are also sensitive to top polarisation effects. 

More generally, it has been shown that the full set of eight $W$ boson spin observables can be measured from selected distributions in top quark decays~\cite{Aguilar-Saavedra:2015yza}, and preliminary measurements have been performed by the ATLAS Collaboration~\cite{ATLAS:2016dcn}.
Moreover, the large top samples available at the LHC Run 1 have allowed to perform more demanding measurements, such as the determination of the two-dimensional $(\theta^*,\phi^*)$ distribution by the ATLAS Collaboration in $t$-channel single top production~\cite{Aad:2015yem}. With the increased statistics at Run 2, it is likely that measurements of the full four-angle distribution, until recently unconceivable, will be achieved. The aim of this work is to provide the framework for such measurements and point out their advantages for a global fit of the $tbW$ vertex. We provide analytical expressions for the fully differential distribution, show how to extract the relevant information either directly or using suitable asymmetries, and relate the top decay amplitudes with $W$ spin observables. Finally, we write the physical quantities involved in the top decay for a general $tbW$ effective Lagrangian, showing that the untangled measurement of the two components of the fraction of longitudinal $W$ bosons may improve the precision of the global fit.

\section{The fully differential distribution}

Using the helicity formalism of Jacob and Wick~\cite{Jacob:1959at}, the amplitude for the top decay $t \to W b \to \ell \nu b$ can be written, in the narrow width approximation, as
\begin{equation}
A_{M \lambda_2 \lambda_3 \lambda_4} = \sum_{\lambda_1} a_{\lambda_1 \lambda_2} b_{\lambda_3 \lambda_4} D_{M \Lambda}^{1/2\,*}(\phi,\theta,0) \, D_{\lambda_1 \lambda}^{1\,*}(\phi^*,\theta^*,0) \,,
\label{ec:A}
\end{equation}
with $\lambda_1$, $\lambda_2$, $\lambda_3$ and $\lambda_4$ the helicities of the $W$ boson, $b$ quark, charged lepton and neutrino, respectively, $M$ the third spin component of the top quark, and $\Lambda = \lambda_1 - \lambda_2$, $\lambda = \lambda_3 - \lambda_4$. The angular dependence is given by the well-known Wigner $D$ functions~\cite{wigner}
\begin{equation}
D^j_{m'm}(\alpha,\beta,\gamma) \equiv \langle jm' | e^{-i \alpha J_z} e^{-i \beta J_y} e^{-i \gamma J_z} | jm \rangle \,,
\end{equation}
and  $a_{\lambda_1 \lambda_2}$, $b_{\lambda_3 \lambda_4}$ are constants. For the top decay, angular momentum conservation implies that there are only four non-zero reduced amplitudes (or just ``amplitudes'' for short) $a_{1 \, \oh}$, $a_{0 \, \oh}$, $a_{0 \, \moh}$ and $a_{-1 \, \moh}$.\footnote{The amplitudes for top quark and anti-quark decays are not equal, but for ease in the notation we denote them with the same symbols, until section~\ref{sec:4} where we write explicit expressions for them.}
 The SM interaction of the $W$ boson with charged leptons implies  $\lambda_3 = \pm 1/2$, $\lambda_4 = \mp 1/2$ for $W^\pm$ decays, assuming massless charged leptons. Therefore, $\lambda = 1$ for top quarks, $\lambda = -1$ for top anti-quarks, and there is only one non-zero $b$ constant in each case, which can be factored out and does not play any further role in the distributions. Let us introduce a top spin density matrix 
\begin{equation}
\rho = \frac{1}{2} \left( \! \begin{array}{cc} 1+P_z & P_x - i P_y  \\ P_x + i P_y & 1-P_z \end{array} \! \right) \,,
\label{ec:rho}
\end{equation}
in terms of the top (anti-)quark polarisation in the three axes, $P_i = 2 \langle S_i \rangle$, with $S_i$ the spin operators. The normalised differential decay width can be written, summing over $b$ quark helicities, as
\begin{eqnarray}
\frac{1}{\Gamma} \frac{d\Gamma}{d\Omega d\Omega^*} & = & \frac{3}{8\pi^2} \frac{1}{\mathcal{N}} \sum_{M M' \lambda_1 \lambda_1' \lambda_2} \rho_{MM'} a_{\lambda_1 \lambda_2} a_{\lambda_1' \lambda_2}^* D_{M\Lambda}^{1/2*} (\phi,\theta,0) D_{M'\Lambda'}^{1/2} (\phi,\theta,0)  \notag \\
& & \times D_{\lambda_1 \lambda}^{1*} (\phi^*,\theta^*,0)  D_{\lambda_1' \lambda}^{1} (\phi^*,\theta^*,0) \,,
\label{ec:fulldist}
\end{eqnarray}
with $d\Omega = d\phi d\!\cos \theta$, $d\Omega^* = d\phi^* d\!\cos \theta^*$, $\Lambda' = \lambda_1' - \lambda_2$ and
\begin{equation}
\mathcal{N} = |a_{1 \, \oh}|^2 + |a_{0 \, \oh}|^2 + |a_{0 \, \moh}|^2 + |a_{-1 \, \moh}|^2 
\end{equation}
the sum of the four non-vanishing amplitudes modulo squared. 
There are only five independent real parameters determining the top decay distribution.
One can define four untangled helicity fractions,
\begin{align*}
F_+^+ = |a_{1,\frac{1}{2}}|^2/\mathcal{N}\,, &\quad F_-^- = |a_{-1,-\frac{1}{2}}|^2/\cal{N}\,,\\
F_0^+ = |a_{0,\frac{1}{2}}|^2/\mathcal{N}\,, &\quad  F_0^- = |a_{0,-\frac{1}{2}}|^2/\cal{N}\,,
\end{align*} 
where $F_+^+$ and $F_-^-$ equal the usual helicity fractions $F_+$ and $F_-$, that is, the relative fractions of $W$ bosons with helicity
$\lambda_1=1$ or $\lambda_1=-1$, respectively. We will thus drop the superscript for these in the rest of the paper. The fraction of $W$ bosons with helicity
$\lambda_1=0$ is $F_0=F_0^+ + F_0^-$. The fractions $F_+^-$  and $F_-^+$, which would correspond to the amplitudes $a_{1 \, \moh}$ and $a_{\mo \, \oh}$, vanish at leading order (LO) due to angular momentum conservation, and are still extremely small at next-to-leading order~\cite{Fischer:2001gp}. The untangled $W$ helicity fractions
yield three independent parameters because their sum equals unity. The remaining two parameters can be taken as the phases of the only interference terms appearing in the sum (\ref{ec:fulldist}),
\begin{equation}
\delta_+ = \arg a_{1 \, \oh} a_{0 \, \oh}^* \,, \quad \delta_- = \arg a_{-1 \, \moh} a_{0 \, \moh}^* \,.
\end{equation}
The property of the $D$ functions
\begin{equation}
D^j_{m'm}(\alpha,\beta,\gamma)^* = (-1)^{m-m'} D^j_{-m' -m}(\alpha,\beta,\gamma)
\end{equation}
and the compostion rule in terms of Clebsch-Gordan coefficients
\begin{equation}
D^{j_1}_{m'_1 m_1}(\alpha,\beta,\gamma) D^{j_2}_{m'_2 m_2}(\alpha,\beta,\gamma) 
= \sum_{j=|j_1-j_2|}^{j_1+j_2}
\langle j_1 m_1' j_2 m_2' | j m' \rangle \langle j_1 m_1 j_2 m_2 | j m \rangle 
D^{j}_{m' m}(\alpha,\beta,\gamma)
\end{equation}
guarantee that the four-dimensional distribution can be expanded as a finite combination of the set of functions we define as
\begin{equation}
M^{j_1 j_2}_{m' m} (\phi,\theta,\phi^*,\theta^*)= \frac{1}{4\pi} ( 2 j_1+1)^{1/2} (2 j_2+1)^{1/2} D_{m'm}^{j_1} (\phi,\theta,0) D_{m0}^{j_2}(\phi^*,\theta^*,0) \,.
\label{ec:M}
\end{equation}
As it is shown in appendix~\ref{sec:a}, these functions are orthonormal. By writing the distribution as
\begin{equation}
\frac{1}{\Gamma} \frac{d\Gamma}{d\Omega d\Omega^*} = \sum_{j_1 j_2 m' m} c^{j_1 j_2}_{m'm} M^{j_1 j_2}_{m'm} \,,
\label{ec:exp}
\end{equation}
we find that the non-zero coefficients in the expansion are
\begin{align}
& c^{00}_{00} = \frac{1}{4\pi} \,, \notag \\
& c^{10}_{00} = \frac{1}{4 \sqrt 3 \pi} P_z \left[ |a_{1 \, \oh}|^2 - |a_{0 \, \oh}|^2 + |a_{0 \, \moh}|^2 - |a_{\mo \, \moh}|^2 \right] / \mathcal{N} \,, \notag \\
& c^{10}_{10} = - (c^{10}_{-1 0})^*  = - \frac{1}{4 \sqrt 6 \pi} (P_x + i P_y) \left[   |a_{1 \, \oh}|^2 - |a_{0 \, \oh}|^2 + |a_{0 \, \moh}|^2 - |a_{\mo \, \moh}|^2 \right] / \mathcal{N} \,, \notag \\
& c^{01}_{00} = \lambda \frac{\sqrt 3}{8 \pi} \left[ |a_{1 \, \oh}|^2 - |a_{\mo \, \moh}|^2
\right] / \mathcal{N} \,, \notag \\
& c^{11}_{00} = \lambda \frac{1}{8 \pi} P_z \left[  |a_{1 \, \oh}|^2 + |a_{\mo \, \moh}|^2
\right] / \mathcal{N} \,, \notag \\
& c^{11}_{10} = - (c^{11}_{-1 0})^* = - \lambda \frac{1}{8 \sqrt 2 \pi} (P_x + i P_y)  \left[  |a_{1 \, \oh}|^2 + |a_{\mo \, \moh}|^2 \right] / \mathcal{N} \,, \notag \\
& c^{11}_{01} = (c^{11}_{0-1})^* =  \lambda \frac{1}{4 \sqrt 2 \pi} P_z \left[  a_{0 \, \oh} \, a_{1 \, \oh}^* + a_{\mo \, \moh} \, a_{0 \, \moh}^*  \right] / \mathcal{N} \,, \notag \\
& c^{11}_{11} = - (c^{11}_{-1 -1})^* = - \lambda \frac{1}{8\pi} (P_x + i P_y) \left[  a_{0 \, \oh} \, a_{1 \, \oh}^* + a_{\mo \, \moh} \, a_{0 \, \moh}^*  \right] / \mathcal{N} \,, \notag \\
& c^{11}_{1-1} = - (c^{11}_{-1 1})^* = - \lambda \frac{1}{8\pi} (P_x + i P_y)  \left[  a_{1 \, \oh} \, a_{0 \, \oh}^*   + a_{0 \, \moh} a_{\mo \, \moh}^*   \right] / \mathcal{N} \,, \notag \\
& c^{02}_{00} = \frac{1}{8 \sqrt 5 \pi} \left[ |a_{1 \, \oh}|^2 - 2 |a_{0 \, \oh}|^2 -2  |a_{0 \, \moh}|^2 + |a_{\mo \, \moh}|^2 \right] / \mathcal{N} \,, \notag \\   \displaybreak
& c^{12}_{00} = \frac{1}{8 \sqrt{15} \pi} P_z \left[ |a_{1 \, \oh}|^2 + 2 |a_{0 \, \oh}|^2 -2  |a_{0 \, \moh}|^2 - |a_{\mo \, \moh}|^2 \right] / \mathcal{N} \,, \notag \\
& c^{12}_{10} = - (c^{12}_{-1 0})^* = - \frac{1}{8 \sqrt{30} \pi} (P_x + i P_y) \left[ |a_{1 \, \oh}|^2 + 2 |a_{0 \, \oh}|^2 - 2  |a_{0 \, \moh}|^2 - |a_{\mo \, \moh}|^2 \right] / \mathcal{N} \,, \notag \\
& c^{12}_{01} = (c^{12}_{0 -1})^* = \frac{1}{4 \sqrt{10} \pi} P_z \left[  a_{0 \, \oh} \, a_{1 \, \oh}^* - a_{\mo \, \moh} \, a_{0 \, \moh}^*  \right] / \mathcal{N} \,, \notag \\ 
& c^{12}_{11} = - (c^{12}_{-1 -1})^* = - \frac{1}{8 \sqrt 5 \pi} (P_x + i P_y) \left[  a_{0 \, \oh} \, a_{1 \, \oh}^* - a_{\mo \, \moh} \, a_{0 \, \moh}^*  \right] / \mathcal{N} \,, \notag \\
& c^{12}_{1-1} = - (c^{12}_{-1 1})^* = - \frac{1}{8 \sqrt 5 \pi} (P_x + i P_y) \left[   a_{1 \, \oh} \, a_{0 \, \oh}^* - a_{0 \, \moh} \, a_{\mo \, \moh}^*   \right] / \mathcal{N} \,.
\label{ec:c}
\end{align}
Because of the orthonormality of the $M$ functions, the different coefficients can be determined by projecting the differential distribution
 $g(\phi,\theta,\phi^*,\theta^*) \equiv (1/\Gamma) d\Gamma/d\Omega d\Omega^*$,
 \begin{equation}
c^{j_1 j_2}_{m'm} = \int d\Omega d\Omega^* g(\phi,\theta,\phi^*,\theta^*) M^{j_1 j_2}_{m'm}(\phi,\theta,\phi^*,\theta^*)^* \,.
\label{ec:proy}
\end{equation}
The Monte Carlo estimate of the integral is the average of the function $(M^{j_1 j_2}_{m'm})^*$ over a selected set of points, given by a probability density function $g$. Therefore, with real data, the estimate of the coefficients $c^{j_1 j_2}_{m'm}$ is simply done by computing the average of $(M^{j_1 j_2}_{m'm})^*$ over the dataset~\cite{Boudreau:2016pdi}, properly corrected for detector effects~\cite{Boudreau:2013yna}.  The statistical error on the coefficients, and the covariance matrix
between different coefficients, can be easily estimated simply by taking averages of the product of $M$ functions.  

\section{Physics parameters from asymmetries}

The set of coefficients $c^{j_1 j_2}_{m'm}$ allows to extract the three top polarisation components and the five physical parameters in the decay of the top quark at the same time. Alternatively, one can measure them using asymmetries or one-dimensional distributions. Using the explicit expressions for the $D$ functions, one can compactly write the fully differential distribution in terms of the four angles as
\begin{eqnarray}
\frac{1}{\Gamma} \frac{d\Gamma}{d\Omega d\Omega^*} & = & \frac{3}{64\pi^2} \frac{1}{\mathcal{N}}
\left\{   \left[ |a_{1 \,\oh}|^2 \left( 1+\lambda \cos \theta^* \right)^2
+ 2 |a_{0 \, \moh}|^2 \sin^2 \theta^*  \right] \left( 1 + \vec P \cdot \vec u_L \right) \right. \notag \\[1mm]
& & + \left[ 2 |a_{0 \, \oh}|^2 \sin^2 \theta^* 
+ |a_{-1 \, -\oh}|^2 \left( 1-\lambda \cos \theta^* \right)^2 \right] \left( 1 - \vec P \cdot \vec u_L \right) \notag \\
& & + \lambda 2 \sqrt 2 \left[ \RE (a_{0 \, \oh} a_{1 \, \oh}^* e^{-i \phi^*}) ( 1+\lambda \cos \theta^*) \right. \notag \\
& & \left. + \RE (a_{-1 \, -\oh} a_{0 \, -\oh}^* e^{-i \phi^*}) (1-\lambda \cos \theta^*)
 \right] \sin \theta^* \vec P \cdot \vec u_T \notag \\
& & + \lambda 2 \sqrt 2  \left[ \IM (a_{0 \, \oh} a_{1 \, \oh}^* e^{-i \phi^*}) (1+\lambda \cos \theta^*) \right. \notag \\
& & + \left. \left. \IM (a_{-1 \, -\oh} a_{0 \, -\oh}^* e^{-i \phi^*}) (1-\lambda \cos \theta^*) 
\right] \sin \theta^* \vec P \cdot \vec u_N \right\} \,.
\end{eqnarray}
Here, $\vec u_L = (\sin \theta \cos \phi,\sin \theta \sin \phi, \cos \theta)$ is the unit vector in the direction of the $W$ boson momentum in the top quark rest frame, and $\vec u_T = (\cos \theta \cos \phi,\cos \theta \sin \phi,-\sin \theta)$, $\vec u_N = (\sin \phi,-\cos \phi,0)$ are two orthonormal vectors. As before, $\lambda = 1$ for top quarks and $\lambda=-1$ for anti-quarks.
The explicit expression above suggests the observables one has to use to extract the desired quantities. First, as mentioned above, the (entangled) $W$ helicity fractions are
\begin{equation}
F_+ = |a_{1 \, \oh}|^2 / \mathcal{N} \,, \quad F_- = |a_{\mo \, \moh}|^2 / \mathcal{N} \,, \quad
F_0 = \left[ |a_{0 \, \oh}|^2 + |a_{0 \, \moh}|^2 \right] /  \mathcal{N} \,,
\label{ec:F}
\end{equation}
and they can be determined from the $\theta^*$ distribution~\cite{Kane:1991bg}. In order to measure the top polarisation, one can use the double forward-backward (FB) asymmetry defined in~\cite{AguilarSaavedra:2012xe} and analogous ones for the $\hat x$ and $\hat y$ axes,
\begin{align}
& A_\text{FB}^{z, z^\prime,} = \frac{1}{\Gamma} \left[ \Gamma(\cos \theta \cos \theta^* > 0) - \Gamma(\cos \theta \cos \theta^* < 0) \right] = \lambda \frac{3}{8} P_z \left[ |a_{1 \, \oh}|^2 +  |a_{\mo \, \moh}|^2 \right] / \mathcal{N} \,, \notag \\
& A_\text{FB}^{x, z'} = \frac{1}{\Gamma} \left[ \Gamma(\cos \phi \cos \theta^* > 0) - \Gamma(\cos \phi \cos \theta^* < 0) \right] = \lambda  \frac{3}{8} P_x \left[ |a_{1 \, \oh}|^2 +  |a_{\mo \, \moh}|^2 \right] / \mathcal{N} \,, \notag \\
& A_\text{FB}^{y, z'} = \frac{1}{\Gamma} \left[ \Gamma(\sin \phi \cos \theta^* > 0) - \Gamma(\sin \phi \cos \theta^* < 0) \right] = \lambda \frac{3}{8} P_y \left[ |a_{1 \, \oh}|^2 +  |a_{\mo \, \moh}|^2 \right] / \mathcal{N} \,,
\end{align}
since the factor between brackets is merely the sum of $F_+$ and $F_-$. The two squared moduli of $\lambda_1=0$ amplitudes, whose sum appears in $F_0$, can be disentangled by a forward-backward-edge-central asymmetry, defined as
\begin{eqnarray}
A_\text{FB,EC}^{z,z'} & = & \frac{1}{\Gamma} \left[ \Gamma(\cos \theta (|\cos \theta^*|-t)  > 0) - \Gamma(\cos \theta (|\cos \theta^*|-t) < 0) \right] \notag \\
& = & \frac{3}{2} (2t-1) P_z \left[ |a_{0 \, \oh}|^2 - |a_{0 \, \moh}|^2 \right] / \mathcal{N} \,,
\end{eqnarray}
with $t=(1+\sqrt 2)^{1/3} - (1+\sqrt 2)^{-1/3} \simeq 0.6$. Alternatively, we point out that the $W$ boson spin analysing power $\alpha_W$ or, equivalently, the FB asymmetry in $\cos \theta$,
\begin{eqnarray}
A_\text{FB}^z & = & \frac{1}{\Gamma} \left[ \Gamma(\cos \theta  > 0) - \Gamma(\cos \theta < 0) \right] \equiv  \frac{1}{2} P_z \alpha_W \notag \\
& = &  \frac{1}{2} P_z \left[ |a_{1 \, \oh}|^2 - |a_{0 \, \oh}|^2 + |a_{0 \, \moh}|^2 -  |a_{\mo \, \moh}|^2 \right] / \mathcal{N} 
\end{eqnarray}
is also sensitive to the difference $|a_{0 \, \oh}|^2 - |a_{0 \, \moh}|^2$. We point out that measuring the untangled helicity fractions $F_0^+$, $F_0^-$ requires polarised top quarks, as it is also seen from the expressions of $c_{m'm}^{j_1 j_2}$. The relative phases of the interfering amplitudes can be extracted from FB and double FB asymmetries,
\begin{eqnarray}
A_\text{FB}^{x'} & = & \frac{1}{\Gamma} \left[ \Gamma(\cos \phi^*  > 0) - \Gamma(\cos \phi^* < 0) \right] =  \frac{3\pi}{8 \sqrt 2} P_z \RE \left[ a_{0 \, \oh}  a_{1 \, \oh}^* + a_{\mo \, \moh} a_{0 \, \moh}^*  \right]  / \mathcal{N} \,, \notag
\end{eqnarray}
\begin{eqnarray}
A_\text{FB}^{y'} & = & \frac{1}{\Gamma} \left[ \Gamma(\sin \phi^*  > 0) - \Gamma(\sin \phi^* < 0) \right] = 
\frac{3\pi}{8 \sqrt 2} P_z \IM \left[ a_{0 \, \oh}  a_{1 \, \oh}^* +  a_{\mo \, \moh}  a_{0 \, \moh}^*  \right]  / \mathcal{N} \,, \notag \\
A_\text{FB}^{x', z' } & = & \frac{1}{\Gamma} \left[ \Gamma(\cos \phi^* \cos \theta^* > 0) - \Gamma(\cos \phi^* \cos \theta^* < 0) \right] \notag \\
& = & \lambda \frac{1}{2 \sqrt 2} P_z \RE \left[ a_{0 \, \oh}  a_{1 \, \oh}^* - a_{\mo \, \moh}  a_{0 \, \moh}^*  \right]  / \mathcal{N} \,, \notag \\
A_\text{FB}^{y', z' } & = & \frac{1}{\Gamma} \left[ \Gamma(\sin \phi^* \cos \theta^* > 0) - \Gamma(\sin \phi^* \cos \theta^* < 0) \right] \notag \\
& = & \lambda \frac{1}{2 \sqrt 2} P_z \IM \left[ a_{0 \, \oh}  a_{1 \, \oh}^* -  a_{\mo \, \moh} a_{0 \, \moh}^* \right]  / \mathcal{N} \,.
\label{ec:asym4}
\end{eqnarray}
As it is apparent from the above equations, the introduction of a FB asymmetry in $\cos \theta^*$ allows to flip the sign of the latter terms between brackets, allowing to measure the real and imaginary parts of each product independently. Again, polarised top quarks are required to measure the quantities, as otherwise there is no privileged direction in the $W$ boson rest frame other than the $\hat z'$ axis.
For completeness, we also give the relation between the eight $W$ spin observables~\cite{Aguilar-Saavedra:2015yza} and top decay amplitudes. They are
\begin{align}
& \langle S_1 \rangle = \lambda \frac{4}{3} A_\text{FB}^{x'} =  \lambda \frac{\pi}{2\sqrt 2} P_z \RE \left[  a_{0 \, \oh}  a_{1 \, \oh}^* +  a_{\mo \, \moh} a_{0 \, \moh}^* \right] / \mathcal{N} \,, \notag \\
& \langle S_2 \rangle =  \lambda \frac{4}{3} A_\text{FB}^{y'} = \lambda \frac{\pi}{2\sqrt 2} P_z \IM \left[  a_{0 \, \oh}  a_{1 \, \oh}^* +  a_{\mo \, \moh} a_{0 \, \moh}^* \right] / \mathcal{N} \,, \notag \\
& \langle S_3 \rangle = F_+ - F_- = \left[ |a_{1 \,\oh}|^2 - |a_{\mo \,\moh}|^2 \right] / \mathcal{N} \,, \notag \\
& \langle T_0 \rangle = \frac{1}{\sqrt 6} \left[  |a_{1 \,\oh}|^2 - 2 |a_{0 \,\oh}|^2 - 2 |a_{0 \,\moh}|^2 + |a_{\mo \,\moh}|^2 \right] / \mathcal{N} \,, \notag \\
& \langle A_1 \rangle = - \frac{\pi}{2} A_\text{FB}^{x' z'} =  - \lambda \frac{\pi}{4 \sqrt 2} P_z \RE \left[  a_{0 \, \oh}  a_{1 \, \oh}^* -  a_{\mo \, \moh} a_{0 \, \moh}^* \right] / \mathcal{N} \,, \notag \\
& \langle A_2 \rangle =  - \frac{\pi}{2} A_\text{FB}^{y' z'} = - \lambda \frac{\pi}{4 \sqrt 2} P_z \IM \left[  a_{0 \, \oh}  a_{1 \, \oh}^* -  a_{\mo \, \moh} a_{0 \, \moh}^* \right] / \mathcal{N} \,,
\end{align}
with $\langle B_1 \rangle = \langle B_2 \rangle = 0$ due to angular momentum conservation.

Finally, let us stress that the global approach in Eqs.~(\ref{ec:exp})--(\ref{ec:proy}) and the use of selected observables (\ref{ec:F})--(\ref{ec:asym4}) are formally equivalent. In particular, some of the asymmetries have a direct relation to coefficients in the expansion, 
\begin{align}
& A_\text{FB}^{z,z'} = 3\pi c_{00}^{11} \,, \quad 
A_\text{FB}^{x,z'} = - 3\sqrt 2 \pi \RE c_{10}^{11} \,, \quad 
A_\text{FB}^{x,z'} = - 3\sqrt 2 \pi \IM c_{10}^{11} \,, \notag \\
& A_\text{FB}^{z} = 2 \sqrt 3 \pi c_{00}^{10} \,, \quad 
A_\text{FB}^{x'} = \lambda  \frac{3 \pi^2}{2} \RE c_{01}^{11} \,, \quad 
A_\text{FB}^{y'} = \lambda  \frac{3 \pi^2}{2} \IM c_{01}^{11} \,, \notag \\
& A_\text{FB}^{x' z'} =  \lambda 2 \sqrt 5 \pi \RE c_{01}^{12} \,, \quad 
A_\text{FB}^{y' z'} =  \lambda 2 \sqrt 5 \pi \IM c_{01}^{12} \,.
\end{align}
However, the calculation of the correlation between observables, which is necessary to include all of them in a global fit, seems easier with the global approach. Whether one method or the other give more precise results, has to be determined with an analysis including all systematic uncertainties.

\section{Physics parameters and the $tbW$ interaction}
\label{sec:4}

The measurement of the top decay amplitudes $a_{\lambda_1 \lambda_2}$ can be interpreted in terms of limits on anomalous $tbW$ interactions. The most general effective $tbW$ interaction arising from the addition of dimension-six operators to the SM Lagrangian can be parameterised as~\cite{AguilarSaavedra:2008zc}
\begin{eqnarray}
\mathcal{L}_{Wtb} & = & - \frac{g}{\sqrt 2} \bar b \, \gamma^{\mu} \left( V_L
P_L + V_R P_R \right) t\; W_\mu^- \nonumber \\
& & - \frac{g}{\sqrt 2} \bar b \, \frac{i \sigma^{\mu \nu} q_\nu}{M_W}
\left( g_L P_L + g_R P_R \right) t\; W_\mu^- + \mathrm{h.c.} \,,
\label{ec:lagr}
\end{eqnarray}
using standard notation, with $g$ the electroweak coupling, $M_W$ the $W$ boson mass and $q_\nu$ its four-momentum. In the SM, $V_L$ equals the Cabibbo-Kobayashi-Maskawa matrix element $V_{tb} \simeq 1$, and the rest of couplings $V_R$, $g_L$ and $g_R$ vanish at the tree level.  For this general vertex, expressions of the $W$ boson spin density matrix have been obtained in Ref.~\cite{AguilarSaavedra:2010nx}. Matching our general expressions for the top decay amplitudes, obtained by general angular momentum conservation arguments, with the explicit calculations there, we obtain for top quarks
\begin{align}
& |a_{1 \, \oh}|^2 = \mathcal{B}_0 + 2 \frac{|\vec q|}{m_t} \mathcal{B}_1 \,, \quad \quad
   |a_{\mo \, \moh}|^2 = \mathcal{B}_0 - 2 \frac{|\vec q|}{m_t} \mathcal{B}_1 \,, \notag \\ 
& |a_{0 \, \oh}|^2 = \frac{1}{2} \mathcal{A}_0 - \frac{|\vec q|}{m_t} \mathcal{A}_1 \,, \quad \quad 
   |a_{0 \, \moh}|^2 = \frac{1}{2} \mathcal{A}_0 + \frac{|\vec q|}{m_t} \mathcal{A}_1 \,, \notag \\
& a_{0 \, \oh} a_{1 \, \oh}^* = \frac{m_t}{\sqrt 2 M_W} (\mathcal{C}_0 - i \mathcal{D}_0) +  \frac{|\vec q|}{\sqrt 2 M_W} (\mathcal{C}_1 - i \mathcal{D}_1) \,, \notag \\
& a_{0 \, \moh} a_{\mo \, \moh}^* = \frac{m_t}{\sqrt 2 M_W} (\mathcal{C}_0 - i \mathcal{D}_0) -  \frac{|\vec q|}{\sqrt 2 M_W} (\mathcal{C}_1 - i \mathcal{D}_1) \,,
\end{align}
up to a global normalisation factor that is irrelevant. Here, $m_t$ is the top quark mass and $|\vec q|$ the modulus of the $W$ boson three-momentum in the top quark rest frame. The form factors $\mathcal{A}_{0,1}$, $\mathcal{B}_{0,1}$, $\mathcal{C}_{0,1}$ and $\mathcal{D}_{0,1}$ depend on the couplings in (\ref{ec:lagr}) and are given in appendix~\ref{sec:b} for completeness. For top anti-quarks the decay amplitudes (denoted here by a bar) are related to the top quark ones by
\begin{align}
& | \bar a_{1 \, \oh}|^2 = |a_{\mo \, \moh}|^2 \,, \quad \quad  |\bar a_{\mo \, \moh}|^2 = |a_{1 \, \oh}|^2 \,, \quad \quad |\bar a_{0 \, \oh}|^2 = |a_{0 \, \moh}|^2 \,, \quad \quad |\bar a_{0 \, \moh}|^2 = | a_{0 \, \oh}|^2 \,, \notag \\
& \bar a_{0 \, \oh} \bar a_{1 \, \oh}^* = \left( a_{0 \, \moh} a_{\mo \, \moh}^* \right)^* \,, \quad \quad
\bar a_{0 \, \moh} \bar a_{\mo \, \moh}^* = \left( a_{0 \, \oh} a_{1 \, \oh}^* \right)^* \,.
\end{align}

\begin{figure}[t]
\begin{center}
\begin{tabular}{ccc}
\includegraphics[height=4.65cm,clip=]{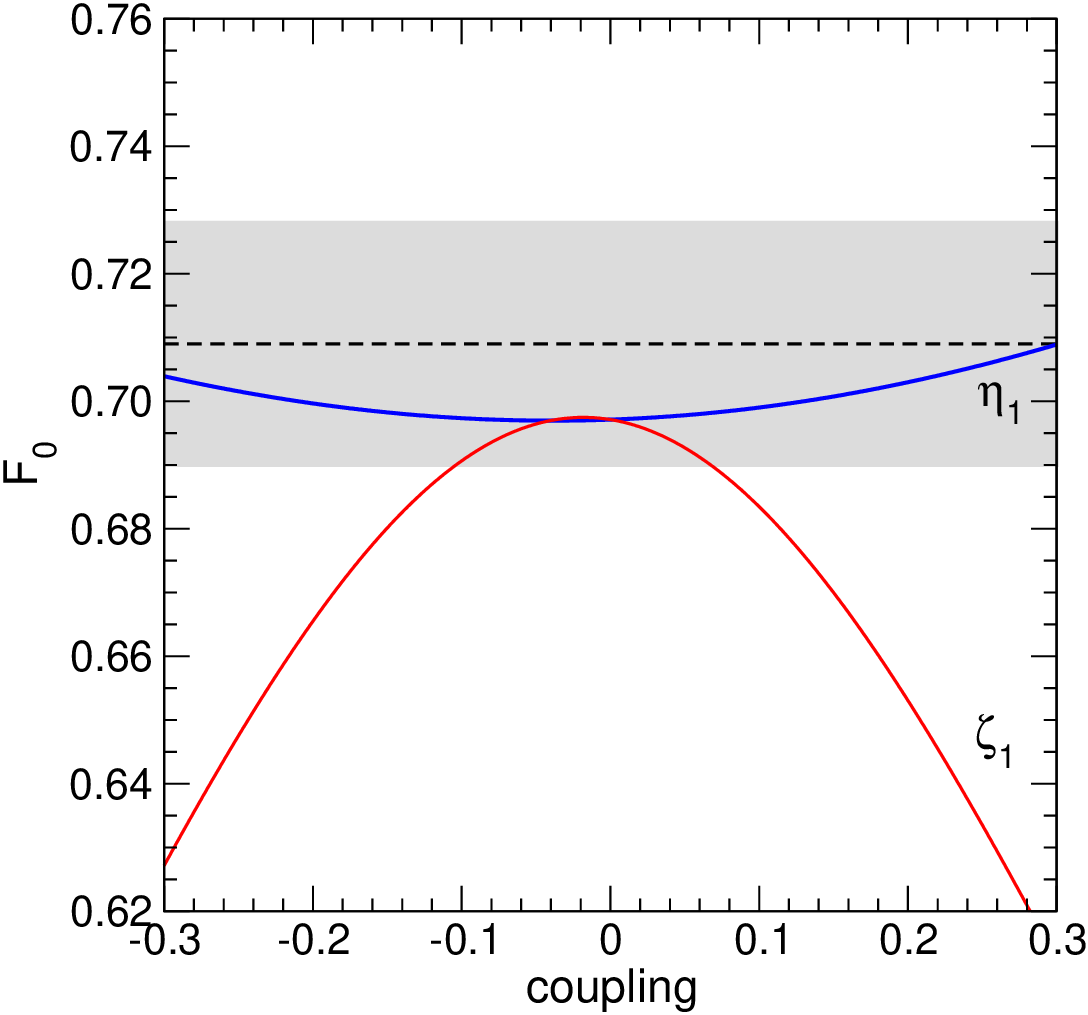} & 
\includegraphics[height=4.65cm,clip=]{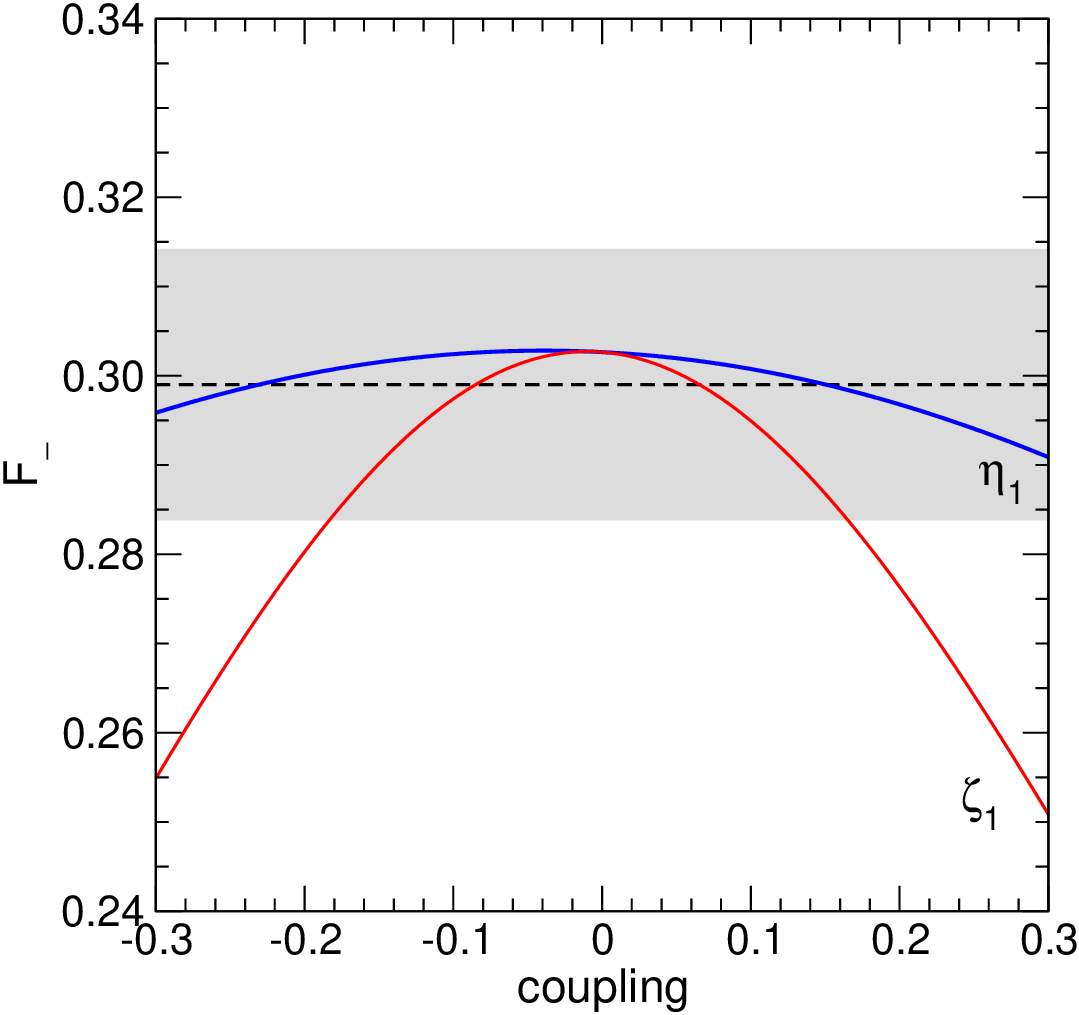} & 
\includegraphics[height=4.65cm,clip=]{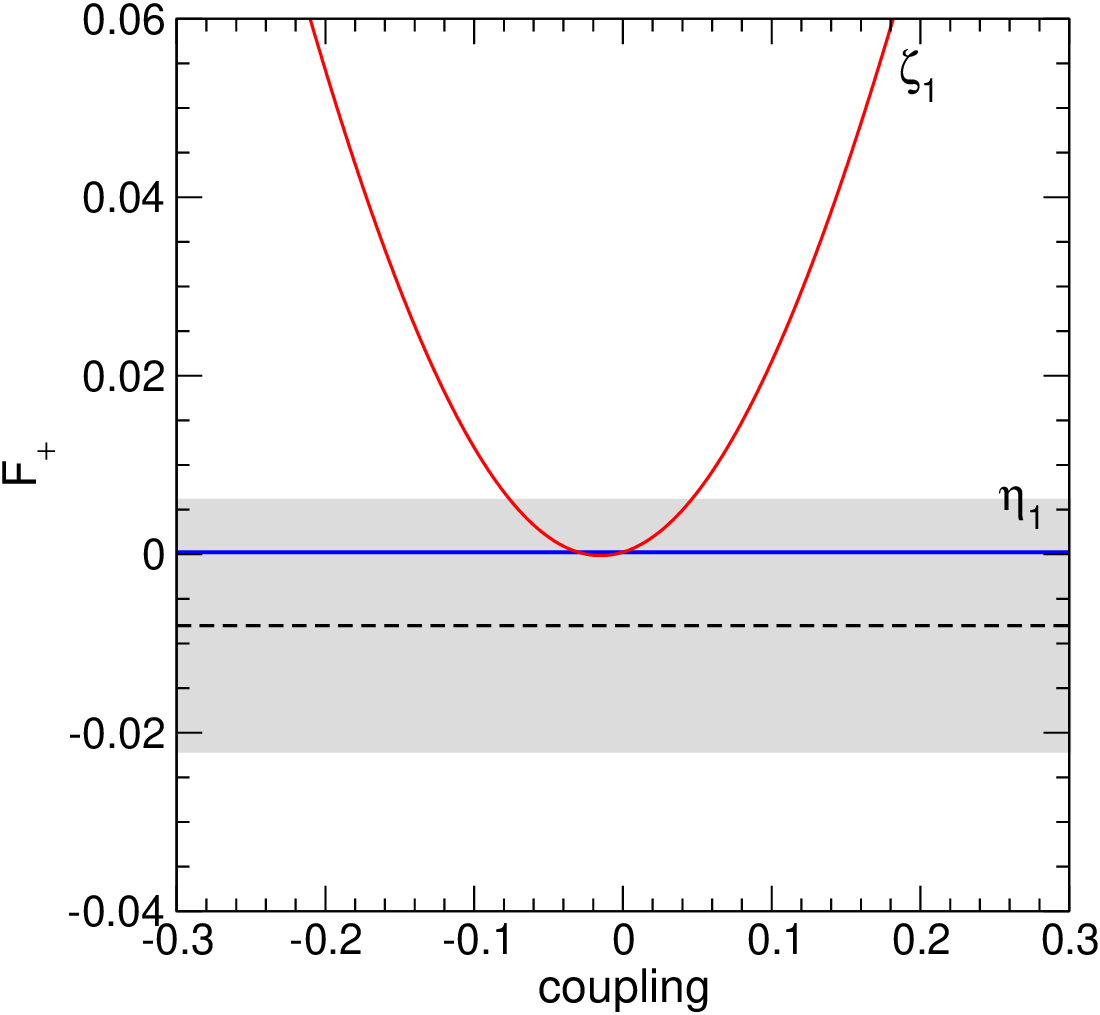}  
\end{tabular}
\end{center}
\caption{Dependence of the helicity fractions on the parameters $\eta_1$ and $\zeta_1$ defined in Eqs.~(\ref{ec:petazeta}). The dashed lines and shaded band represent the current most precise measurements~\cite{Aaboud:2016hsq} and their uncertainty.}
\label{fig:F}
\end{figure}

\begin{figure}[t]
\begin{center}
\includegraphics[height=4.8cm,clip=]{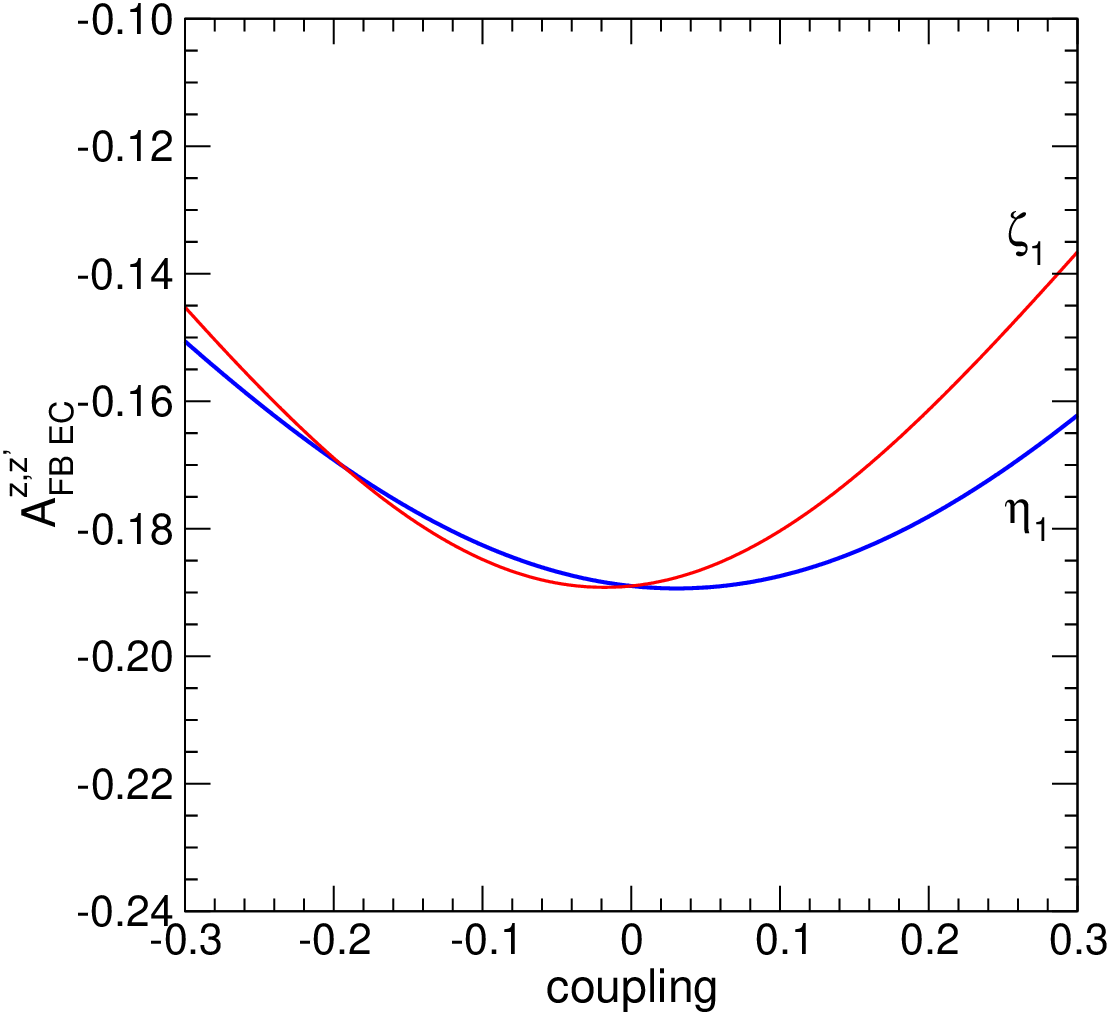}
\end{center}
\caption{Dependence of the asymmetry $A_\text{FB,EC}^{z,z'}$ on the parameters $\eta_1$ and $\zeta_1$ defined in Eqs.~(\ref{ec:petazeta}).}
\label{fig:AFBEC}
\end{figure}

It is known~\cite{AguilarSaavedra:2008gt} that there is a cancellation between anomalous contributions to helicity fractions when $M_W V_R \simeq m_t g_L$, as well as when $M_W V_L \simeq m_t g_R$. This cancellation stems from the Gordon identities that one can write for on-shell $t$ and $b$ quarks,
\begin{align}
\bar b(p_b) \left[ i \sigma^{\mu \nu} (p_t - p_b)_\nu P_L + m_t \gamma^\mu P_R + m_b \gamma^\mu P_L \right] t(p_t) = \bar b(p_b) (p_t+p_b)^\mu P_L t(p_t) \,, \notag \\
\bar b(p_b) \left[ i \sigma^{\mu \nu} (p_t - p_b)_\nu P_R + m_t \gamma^\mu P_L + m_b \gamma^\mu P_R \right] t(p_t) = \bar b(p_b) (p_t+p_b)^\mu P_R t(p_t) \,.
\end{align}
Neglecting the $b$ quark mass, the combinations of couplings with  $M_W V_R = m_t g_L$ or 
$M_W V_L = m_t g_R$ are equivalent to an interaction of the type $(p_t+p_b)^\mu P_{L,R}$, which does not contribute for any element of the $W$ boson spin density matrix except for $\lambda_1 = \lambda_1' = 0$, because the product of $p_b$ with the $W$ polarisation vectors of helicities $\pm 1$ vanishes.\footnote{Interactions of this type arise from the dimension-six effective operators $O_{Du}^{ij}$, $O_{\bar Du}^{ij}$, $O_{Dd}^{ij}$ and $O_{\bar Dd}^{ij}$, which were shown to be redundant in Ref.~\cite{AguilarSaavedra:2008zc}. Therefore, the insensitivity to these combinations of couplings can be viewed as insensitivity to these effective operators.}
Therefore, the effect in the $W$ spin observables is residual, and given by the change in the partial width to $\lambda_1 = 0$ states. The cancellation is apparent if we define new couplings $\eta_{1,2}$, $\zeta_{1,2}$ as unitary rotations of the ones in (\ref{ec:lagr}),
\begin{align}
& \left(\! \begin{array}{c} V_R \\ g_L \end{array} \! \right)
= \frac{1}{(m_t^2 + M_W^2)^{1/2}}
\left( \! \begin{array}{cc} m_t & -M_W \\ M_W & m_t \end{array} \!  \right)
\left(\! \begin{array}{c} \eta_1 \\ \zeta_1 \end{array} \! \right)
\,, \quad \notag \\[2mm]
& \left(\! \begin{array}{c} V_L \\ g_R \end{array} \! \right)
= \frac{1}{(m_t^2 + M_W^2)^{1/2}}
\left( \! \begin{array}{cc} m_t & -M_W \\ M_W & m_t \end{array} \!  \right)
\left(\! \begin{array}{c} \eta_2 \\ \zeta_2 \end{array} \! \right)
\,.
\label{ec:petazeta}
\end{align}
The dependence of the helicity fractions on $\eta_1$ and $\zeta_1$, for $V_L = 1$ and $g_R = 0$, is depicted in Fig.~\ref{fig:F}, at LO. (Next-to-next-to-leading order calculations of the helicity fractions are available~\cite{Czarnecki:2010gb}, but since the difference with LO is smaller than the experimental uncertainty, we use the latter for consistency.) The helicity fractions are rather insensitive to $\eta_1$, or, in other words, a combination with $\zeta_1 = 0$ ($M_W V_R  = m_t g_L$) gives a very small effect. The remaining $W$ boson spin observables exhibit the same behaviour.
This can also be seen analytically. Neglecting the $b$ quark mass, the form factors in appendix~\ref{sec:b} read
\begin{align}
& \mathcal{A}_0 = \frac{m_t^2 - M_W^2}{m_t^2 M_W^2 (m_t^2 + M_W^2)} 
\left| (m_t^2 - M_W^2) \eta_1 - 2 m_t M_W \zeta_1 \right|^2 + (1 \to 2) \,, \notag \\
& \mathcal{A}_1 = - \frac{1}{M_W^2 (m_t^2 + M_W^2)} \left| (m_t^2 - M_W^2) \eta_1 - 2 m_t M_W \zeta_1 \right|^2 - (1 \to 2) \,, \notag \\ 
& \mathcal{B}_0 = \frac{m_t^4 - M_W^4}{m_t^2 M_W^2} \left| \zeta_1 \right|^2 + (1 \to 2) \,, \notag \\
& \mathcal{B}_1 = \frac{m_t^2 + M_W^2}{M_W^2} \left| \zeta_1 \right|^2 - (1 \to 2) \,, \notag \\
& \mathcal{C}_0 = \frac{m_t^2 - M_W^2}{m_t^2} \left[ 2 \left| \zeta_1 \right|^2 - \frac{m_t^2 - M_W^2}{m_t M_W} \RE \zeta_1 \eta_1^* \right]  + (1 \to 2) \,, \notag \\
& \mathcal{C}_1 = 2 \left[ 2 \left| \zeta_1 \right|^2 - \frac{m_t^2 - M_W^2}{m_t M_W} \RE \zeta_1 \eta_1^* \right]  - (1 \to 2) \,, \notag \\
& \mathcal{D}_0 = - \frac{(m_t^2 - M_W^2)^2}{m_t^3 M_W} \IM \zeta_1 \eta_1^* + (1 \to 2) \,, \notag \\
& \mathcal{D}_1 = - 2 \frac{m_t^2 - M_W^2}{m_t M_W}  \IM \zeta_1 \eta_1^* - (1 \to 2) \,.
\end{align}
We observe that for $\zeta_1 = 0$, the anomalous contributions from $\eta_1$ to the $\mathcal{B}_{0,1}$, $\mathcal{C}_{0,1}$ and $\mathcal{D}_{0,1}$ form factors vanish, in agreement with our previous argument, but for $\mathcal{A}_0$ and $\mathcal{A}_1$, precisely the form factors involved in $F_0^+$ and $F_0^-$, they do not. Therefore, the measurement of these two untangled helicity fractions, for example using $A_\text{FB,EC}^{z,z'}$, can break the degeneracy. For illustration, we plot in Fig.~\ref{fig:AFBEC} the dependence of $A_\text{FB,EC}^{z,z'}$ on $\eta_1$ and $\zeta_1$, for $V_L = 1$ and $g_R = 0$. There is a sharp difference with the helicity fractions, and for this asymmetry the variations with $\eta_1$ and $\zeta_1$ are similar. The same information can of course be obtained, in the global analysis, by the measurement of $c_{00}^{10}$ and $c_{00}^{12}$.

\section{Summary}

In this work we have set the framework to extract all the relevant physical quantities from the measurement of the four-dimensional top decay distribution, that is, the top polarisation in three orthogonal directions and the five parameters determining the top decay distributions. Such demanding measurements may be possible in the near future with the LHC Run 2 data. With the full distribution, or with a suitable forward-backward-edge-central asymmetry, it will be possible to measure the two $\lambda_1= 0$ untangled  $W$ boson helicity fractions $F_0^+$ and $F_0^-$. The precision of the global fit to the $tbW$ vertex might be substantially improved if these untangled helicity fractions are accurately measured in upcoming analyses.

\section*{Acknowledgements}
This research has been supported by MINECO Projects  FPA 2016-78220-C3-1-P, FPA2015-65652-C4-1-R, FPA 2013-47836-C3-2-P (including ERDF), Junta de Andaluc\'{\i}a Project FQM-101, Generalitat Valenciana Project PROMETEO/2010/021, European Commission through the contract PITN-GA-2012-316704 (HIGGSTOOLS), and by the U.S. Department of Energy under Grant No. DE-SC0007914.

\appendix
\section{Orthonormality of the $M$ functions}
\label{sec:a}

The functions $D^j_{m' m}(\phi,\theta,0)$ with the third argument set to zero are not orthogonal when integrated with respect to $d\Omega$; rather, orthogonality holds for the functions $D^j_{m' m}(\alpha,\beta,\gamma)$ when integrated over the three angles. However, the $M$ functions defined in (\ref{ec:M}) are orthogonal because the second index in the first $D$ function is precisely the first index in the second $D$ function. Explicitly, we have
\begin{align}
& \int d\Omega d\Omega^* M^{j_1 j_2}_{r s} (M^{j'_1 j'_2}_{r' s'})^* \notag \\
& \quad = \frac{1}{16 \pi^2} \left[ (2j_1+1) (2j'_1 +1) (2j_2+1) (2j'_2+1) \right]^{1/2} \int d\phi d\phi^* e^{i \phi(r'-r)} e^{i \phi^*(s'-s)} \notag \\
& \quad \quad \times \int d\!\cos \theta d\!\cos \theta^* d^{j_1}_{rs}(\theta) d^{j'_1}_{r's' }(\theta) d^{j_2}_{s0}(\theta^*) d^{j'_2}_{s' 0 }(\theta^*) \notag \\
& \quad = \frac{1}{4}  \left[ (2j_1+1) (2j'_1 +1) (2j_2+1) (2j'_2+1) \right]^{1/2} \delta_{rr'} \delta_{ss'} \notag \\
& \quad \quad \times \int d\!\cos \theta d\!\cos \theta^* d^{j_1}_{rs}(\theta) d^{j'_1}_{rs }(\theta) d^{j_2}_{s0}(\theta^*) d^{j_2}_{s 0}(\theta^*) \,,
\end{align} 
with $d^j_{m'm}(\beta) \equiv \langle jm' | e^{-i \beta J_y} | jm \rangle$ the small $d$ Wigner functions, which are real, and satisfy
\begin{equation}
\int d\!\cos \theta d^{j}_{m'm }(\theta) d^{j'}_{m'm}(\theta) = \frac{2}{2j+1} \delta_{jj'} \,.
\label{ec:dorth}
\end{equation}
Using (\ref{ec:dorth}), one easily arrives at 
\begin{align}
& \int d\Omega d\Omega^* M^{j_1 j_2}_{r s} (M^{j'_1 j'_2}_{r' s'})^* = \delta_{j_1 j'_1} \delta_{j_2 j'_2} \delta_{rr'} \delta_{ss'} \,.
\end{align} 

\section{Expressions for top decay form factors}
\label{sec:b}

We collect here the expressions for the eight dimensionless form factors involved in the decay $t \to Wb$ for a general $tbW$ vertex, from Ref.~\cite{AguilarSaavedra:2010nx}. Defining  $x_W = M_W/m_t$, $x_b = m_b/m_t$, with $m_b$ the $b$ quark mass, they are
\begin{align}
\mathcal{A}_0 & = \frac{m_t^2}{M_W^2} \left[ |\vl|^2 + |\vr|^2 \right] \left(1 - x_W^2 \right)
+ \left[ |\gl|^2 + |\gr|^2 \right] \left(1 - x_W^2 \right) \notag \\
&  - 4 x_b \, \RE \left[ \vl \vr^* + \gl \gr^* \right]
- 2 \frac{m_t}{M_W} \RE \, \left[\vl \gr^* + \vr \gl^* \right]
\left(1 - x_W^2 \right)  \notag \\
& + 2 \frac{m_t}{M_W} x_b \,\RE \, \left[\vl \gl^* + \vr \gr^* \right]
\left(1 +x_W^2 \right) \,, \notag \\
\mathcal{A}_1 & = \frac{m_t^2}{M_W^2} \left[ |\vl|^2 - |\vr|^2 \right]
- \left[ |\gl|^2 - |\gr|^2 \right]
- 2 \frac{m_t}{M_W} \RE \,\left[ \vl \gr^* - \vr \gl^* \right] \notag \\
& + 2 \frac{m_t}{M_W} x_b \RE \, \left[ \vl \gl^* - \vr \gr^* \right] \,, \notag \\ 
\mathcal{B}_0 & = \left[ |\vl|^2 + |\vr|^2 \right] \left(1 - x_W^2 \right) + \frac{m_t^2}{M_W^2} \left[ |\gl|^2 + |\gr|^2 \right] \left(1 - x_W^2 
 \right) \notag \\
& - 4 x_b \, \RE \left[ \vl \vr^* + \gl \gr^* \right]
- 2 \frac{m_t}{M_W} \RE \, \left[\vl \gr^* + \vr \gl^* \right]
\left(1 - x_W^2 \right) \notag \\ 
& + 2 \frac{m_t}{M_W} x_b \,\RE \, \left[\vl \gl^* + \vr \gr^* \right]
\left(1 +x_W^2 \right) \,, \notag
 \\ 
\mathcal{B}_1 & = - \left[ |\vl|^2 - |\vr|^2 \right] 
 + \frac{m_t^2}{M_W^2} \left[ |\gl|^2 - |\gr|^2 \right]
+ 2 \frac{m_t}{M_W} \, \RE \, \left[\vl \gr^* - \vr \gl^* \right]  \notag \\
&  + 2 \frac{m_t}{M_W} x_b \, \RE \, \left[\vl \gl^* - \vr \gr^* \right] \,,
\notag \\
\mathcal{C}_0 & = \left[ |\vl|^2 + |\vr|^2 + |\gl|^2 + |\gr|^2  \right] \left(1 - x_W^2 \right) 
- 2 x_b \, \RE \left[ \vl \vr^* + \gl \gr^* \right] \left( 1+x_W^2 \right) \notag \\
& - \frac{m_t}{M_W} \RE \, \left[\vl \gr^* + \vr \gl^* \right]
\left(1 - x_W^4 \right) + 4 x_W x_b \,\RE \, \left[\vl \gl^* + \vr \gr^* \right]
\,, \notag \\
\mathcal{C}_1 & = 2 \left[ -|\vl|^2 + |\vr|^2 + |\gl|^2 - |\gr|^2 \right] + 2 \frac{m_t}{M_W}
\RE \, \left[ \vl \gr^* - \vr \gl^* \right] \left( 1+x_W^2 \right)
\,, \notag \\
\mathcal{D}_0 & = \frac{m_t}{M_W} \IM \, \left[ \vl \gr^* + \vr \gl^* \right] 
\left( 1-2 x_W^2 + x_W^4 \right) \,, \notag \\
\mathcal{D}_1 & = -4 x_b \, \IM \left[ \vl \vr^* + \gl \gr^* \right] -2 \frac{m_t}{M_W}
\IM \left[\vl \gr^*-\vr \gl^* \right] (1-x_W^2) \,.
\label{ec:AtoF}
\end{align}

\end{document}